\documentclass[journal,10pt,onecolumn]{IEEEtran}
\usepackage{amsmath,amsthm}
\usepackage{amsfonts}
\usepackage{amssymb}
\usepackage{mathrsfs}
\usepackage{mathtools}
\usepackage[english]{babel}
\usepackage{float}
\usepackage{amsmath}
\usepackage{color}
\usepackage{multirow}
\usepackage{graphicx}

\usepackage[font=scriptsize,caption=false,labelsep=space]{subfig}
\usepackage{bm}
\usepackage{balance}
\usepackage{cite}
\usepackage{comment}

\numberwithin{equation}{section}
 
\newtheorem{theorem}{Theorem}

\newtheorem{proposition}[theorem]{Proposition}

\usepackage[switch]{lineno}
\usepackage[named]{algo}
\usepackage[noend]{algpseudocode}
\usepackage{algorithm}
\newcommand{\dddag}{%
  \mathbin{\vbox{\offinterlineskip\ialign{%
    \hfil##\hfil\cr
    \small{$\dagger$}\cr
    \noalign{\kern-0.6ex}
    \small{$\ddagger$}\cr
}}}}

\title{Multi-Carrier Wideband OCDM-Based THz Automotive Radar}%
	
\author{Sangeeta Bhattacharjee, Kumar Vijay Mishra, Ramesh Annavajjala and Chandra R. Murthy

\thanks{S. B. and C. R. M. are with the Indian Institute of Science, Bangalore 560012 India, e-mail: \{bsangeeta, cmurthy\}@iisc.ac.in.}
\thanks{K. V. M. is with the United States CCDC Army Research Laboratory, Adelphi, MD 20783 USA, e-mail: kvm@ieee.org.}
\thanks{R. A. is with the University of Massachusetts, Boston, MA 02125 USA, email: ramesh.annavajjala@umb.edu.}
}

\begin{document}

\maketitle

\begin{abstract}
Automotive radars at the Terahertz (THz) frequency band have the potential to be compact and lightweight while providing high (nearly-optical) angular resolution. In this paper, we propose a bistatic THz automotive radar that employs the recently proposed orthogonal chirp division multiplexing (OCDM) multi-carrier waveform. As a stand-alone communications waveform, OCDM has been investigated for robustness against interference in time-frequency selective channels. The THz-band path loss, and, hence, radar signal bandwidth, are range-dependent. We address this unique feature through a multi-carrier wideband OCDM sensing transceiver that exploits the coherence bandwidth of the THz channel. 
We develop an optimal scheme to combine the returns at different range/bandwidths by assigning weights based on the Cram\'{e}r-Rao lower bound on the range and velocity estimates. Numerical experiments 
demonstrate improved target estimates using our proposed combined estimation from multiple varied-attenuation THz frequencies.
\end{abstract}

    \section{Introduction}
\label{sec:intro}
Autonomous driving is one of the mega-trends of automotive industry, wherein the majority of auto makers have already introduced various levels of autonomy into commercially available vehicles \cite{bilik2019rise,sun2020mimo}. While multiple sensors such as camera, radar, lidar, and ultrasonics are used to enable autonomy in vehicles, radar is preferred as an inexpensive, all-weather sensor \cite{patole2017automotive}. However, at present, millimeter-wave automotive radars with a few GHz bandwidth \cite{dokhanchi2019mmwave} at $24$ and $77$ GHz are unable to achieve the high-resolution images of optical sensors \cite{wang2020displaced}. As a result, there is a gradual push to sense the automotive environment at Terahertz (THz) frequency band \cite{norouzian2019next,elbir2021terahertz,xiao2020modeling}. There are multiple advantages of higher frequency operation, such as a smaller size system, feasibility of a larger number of channels leading to higher angular resolution, and availability of higher bandwidth that yields higher range resolution. Currently, \textit{low-THz} frequencies, such as $0.15$ THz and $0.3$ THz bands provide $6$ GHz and $16$ GHz unlicensed contiguous bandwidths, respectively. These wide frequency bands enable automotive radars to achieve lidar-like imaging capabilities  \cite{marchetti2018radar,sheeny2020300,marchetti2019automotive}. 


A major drawback of operating automotive radars above $100$ GHz is the high propagation losses because of increased atmospheric absorption and attenuation 
\cite{jornet2011channel}. Further, the low-THz spectrum exhibits distance-dependent spectral windows. While the entire band may be considered as a single transmission window with a bandwidth of the order of a THz at distances below $1$ meter, there are multiple transmission windows that are tens or hundreds of GHz wide at higher distances because of increased molecular absorption. In fact, the bandwidth of each transmission window shrinks with the transmission distance, and reduces by an order of magnitude when the distance is increased from 1 to 10 meters due to high absorption peaks \cite{hossain2019hierarchical}. Thus, in the THz band, there is a critical trade-off between operating the automotive radar at high bandwidth, thereby improving range resolution, and maintaining an adequate maximum detectable range. 

Several interesting works, focusing on the classical techniques such as MIMO beamforming and precoding \cite{elbir2021terahertz}, transceiver design \cite{han2015multi}, and waveform design \cite{wu2021non, hossain2019hierarchical} have been investigated to tackle high propagation losses and power limitations of THz bands \cite{sarieddeen2021overview}. In \cite{jornet2014femtosecond}, a single-band pulse-based scheme is proposed at the THz frequencies. However, this modulation is valid mainly for very short transmission distances, e.g., in nano networks, where distance-dependent spectral windows do not appear. In \cite{hossain2019hierarchical}, a multi-narrowband system is developed. However, the resulting number of sub-bands is large and the hierarchical modulation is complicated. Moreover, the effect of inter-channel interference is neglected. 

Traditionally, single carrier (SC) radar probing waveforms  
have been explored for sensing \cite{jiang2016effective,li2020non} above 90 GHz because of higher spectral and energy efficiency and lower interference along with reduced implementation complexity compared to multi-carrier waveforms 
\cite{sarieddeen2021overview}. However, frequency selectivity as function of the target distance, number of multipaths, pulse bandwidth, and center frequency can still arise in low-THz
systems because of frequency- and distance-dependent molecular absorption losses. 
In this context, our goal is to investigate the efficacy of an emerging multicarrier waveform -- orthogonal chirp-division multiplexing (OCDM) \cite{de2020ocdm, sang2022integrated} -- for THz automotive sensing. The  motivation of using OCDM as the underlying waveform for radar comes from the fact that it is a spread spectrum technique that employs multiplexing of orthogonal chirp signals as carriers for data transmission. These orthogonal chirp signals are also well-suited for radar applications due to their superior pulse compression characteristics, better robustness against interference and comparable implementation complexity as opposed to conventional orthogonal frequency-division multiplexing (OFDM) \cite{sang2022integrated,ouyang2016orthogonal}. 

We propose a THz radar sensing framework using multi-carrier wideband OCDM (MCW-OCDM) based transceivers, each tuned to center frequencies of available spectral windows that are artefacts of THz band. The target is independently estimated at each of these windows. We develop a multi-stage sensing framework for the MCW-OCDM THz system, which exploits the strength of radar returns for optimally combining  the individual \emph{estimates} at each sensing processor to localize the range and velocity of targets. The proposed system is able to tackle both distance and frequency dependent path losses in the THz band.
Our numerical experiments show that the proposed system realizes sub-millimeter-level accuracy for range estimation, which is a three orders of magnitude improvement, compared to processing the returns from a single transmission window. Furthermore, the velocity estimates are significantly improved over that of an individual transmission window, across varying target distances.
\section{System Model}
\label{sec:system model}
We consider a bistatic automotive system, where the signal transmitted by a transmit (Tx) vehicle is reflected off $P$ targets of interest and then captured by a receive (Rx) vehicle. The radar scene comprises of $P$ non-fluctuating point-targets following the Swerling-0 target model \cite{skolnik2008radar}. We assume that a target position relative to the bistatic radar varies linearly throughout the time-on-target of the Tx signal, i.e., $r_p(t) =r_p+v_pt$, where $r_p$ is the initial range at $t=0$ and $v_p$ is the constant radial velocity. The signal emitted by the Tx passes through a frequency-selective time-varying channel in the THz band with impulse response \cite{han2014multi}
\begin{equation} \label{eq:generic channel}
h(t,\tau)=\sum \limits_{p=0}^{P-1} h_p \exp(\mathrm j 2 \pi f_{D_p} t) \delta (\tau-\tau_p),
\end{equation}
where $h_p$ is the complex scattering coefficient of the p-th point target,  $\tau_{p}=\tau^{(1)}_{p}+\tau^{(2)}_{p}$ is the time delay, which is linearly proportional to the target's bi-static range $r_p = c\tau_p$ (superscripts (1) and (2) denote variable dependency on the Tx-target and target-Rx paths, respectively), $f_{D_p}=f_{D_p}^{(1)}+f_{D_p}^{(2)}$ is the Doppler shift induced by the target's linear motion  $v_p= c\frac{f_{D_p}}{ f_c}$, with $f_c$ being the operating frequency and $c$ denoting the speed of light. Note that the path gain of the received signal power due to reflection from the $p$-th target is characterized by the path loss of the THz band channel
\begin{equation}
\label{eq:path loss}
   \text{ PL}_{\text{LoS}} (f_c,r_p)= \left( \frac{4 \pi f_c r_p }{c} \right)^2 e^{k_{\text{abs}}(f_c) r_p},
\end{equation}
where $k_{\text{abs}}$ is the frequency-dependent absorption coefficient of the medium \cite{han2022terahertz, jornet2011channel}. \par

\begin{figure}[t]
      \centering
    \includegraphics[width=0.6\textwidth]{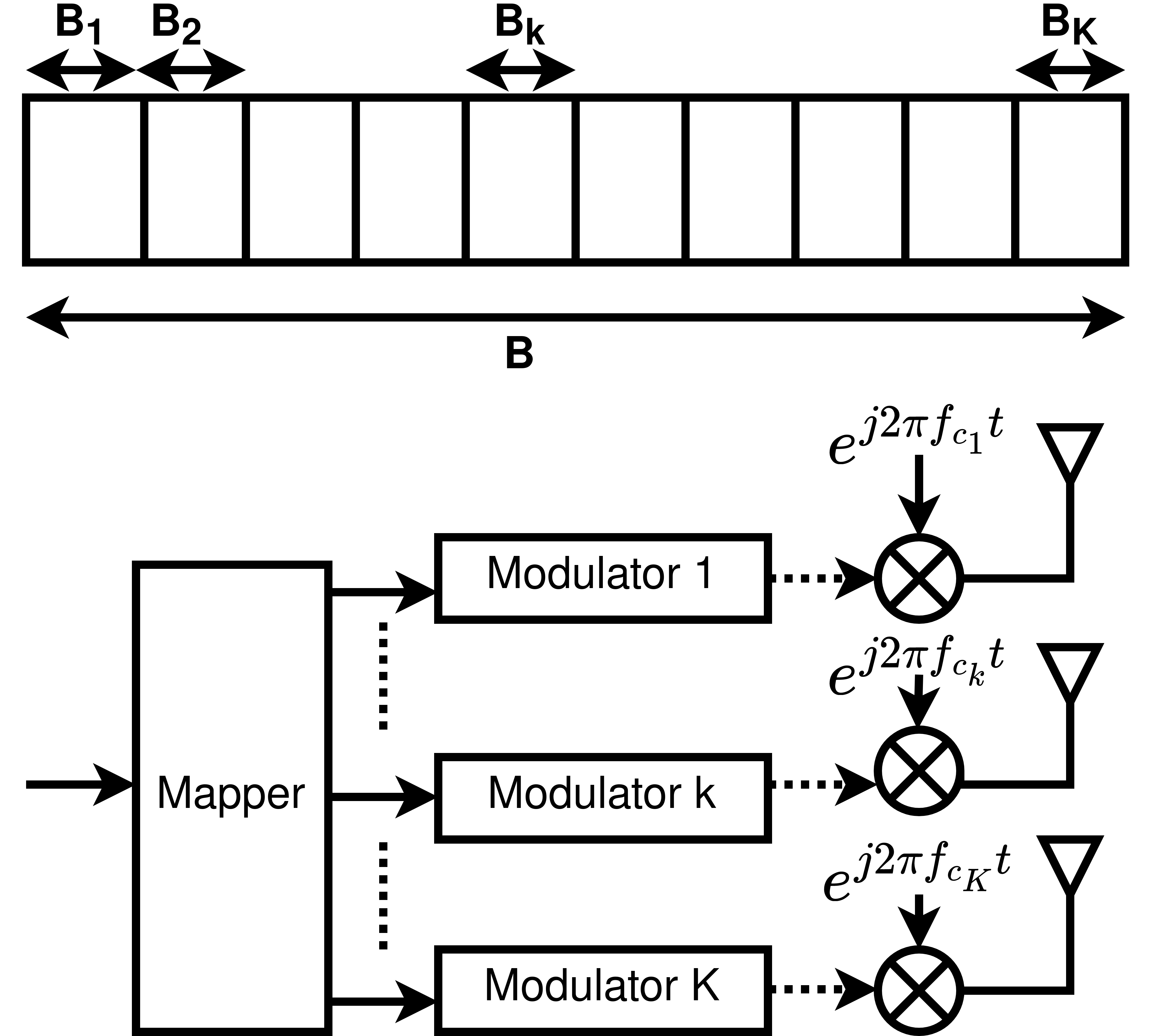}
    \caption{The MCW-OCDM radar Tx multiplexes several chirps in the bandwidth $B$ that is divided into $K$ (not necessarily equal) subbands. The input from payload data frames are mapped into $K$ independent OCDM modulator blocks.}
    \label{fig:Transmitter}
\end{figure}

The Tx  waveform is an OCDM signal, which multiplexes a bank of chirps in the same time-period and bandwidth. The total bandwidth $B$ is divided into $K$ subbands, with the $i$th subband spanning a of bandwidth $B_i$ at center frequency $f_{c_i}$, $i \in \lbrace 1, \cdots , K\rbrace$. The input from payload data frames are mapped into $K$ independent OCDM modulator blocks as shown in Fig. \ref{fig:Transmitter}. The Tx frame in the $i$th subband consists of $N_i$ temporal symbols obtained by modulating the phase and amplitude of $M_i$ sub-chirps using the data bits, and occupying a total bandwidth $B_i=M_i \Delta f_i$, where $\Delta f_i$ is the bandwidth of each chirp and $M_i$ is an even positive integer. With such a chirp basis, the baseband Tx OCDM signal at the $i$th modulator output is \cite{ouyang2016orthogonal}
\begin{align} \label{eq:ocdm symbol}
& \hspace{3cm} \mathbf{S}_i=\mathbf{\Phi}_{M_i}^H \mathbf{X}_i \nonumber \\
&\hspace{-0.3cm}\text{where} \
\mathbf{\Phi}_{M_i}=\frac{1}{\sqrt M_i} \mathbf{\Theta}_1 \mathbf{F}_{M_i} \mathbf{\Theta}_2,
\text{with} \ \ \mathbf{F}_{M_i} = \left[\frac{1}{\sqrt {M_i}}  e^{\mathrm j\frac{2\pi}{M_i}uv}\right],  \\ & \mathbf{\Theta}_1=\text{diag} \lbrace \Theta_{1,0}; \cdots;\Theta_{1,M_i-1}\rbrace, \ \  \Theta_{1,u}=e^{-\mathrm j\frac{\pi}{4}}e^{\mathrm j\frac{\pi}{M_i}u^2}, \nonumber \\ & \mathbf{\Theta}_2=\text{diag} \lbrace \Theta_{2,0}; \cdots;\Theta_{2,M_i-1}\rbrace, \ \  \Theta_{2,v}=e^{\mathrm j\frac{\pi}{M_i}v^2},
\end{align}
where $u, v \in \{1, \cdots, M_i\}$, $\mathbf \Phi_{M_i}^H \in \mathbb{C}^{M_i \times M_i}$ denotes the inverse discrete Fresnel transform (IDFnT) of order $M_i$ and $\mathbf{X}_i \in \mathbb{C}^{M_i \times N_i}$ is the matrix of data symbols, $[\mathbf X_{i}]_{m,n}$, where $m=\lbrace 0, \cdots, M_i-1 \rbrace$ is the chirp index and $n=\lbrace 0, \cdots, N_i-1\rbrace$ is the symbol index. The IDFnT is the product of the discrete Fourier transform (DFT) matrix $\mathbf{F}_{M_i}$ and additional quadratic phases. Furthermore, the matrix $\mathbf{\Phi}_{M_i}$ is circulant. Hence, using the eigen-decomposition property, \eqref{eq:ocdm symbol} becomes
\begin{align} \label{eq:ocdm eigen}
\mathbf{S}_i&=\mathbf{F}_{M_i}^H \mathbf{\Gamma}^H\mathbf{F}_{M_i} \mathbf{X}_i=\mathbf{F}_{M_i}^H\mathbf{Z}_i
\end{align}
where $\mathbf{\Gamma}^H=\mathbf{F}_{M_i}\mathbf{\Phi}_{M_i}^H\mathbf{F}_{M_i}^H$ and the matrix $\mathbf{G}\triangleq \mathbf{\Gamma}^H\mathbf{F}_{M_i} \in \mathbb{C}^{M_i \times M_i}$ transforms the input data symbols $\mathbf{X}_i$ into scaled frequency domain symbols $\mathbf{Z}_i\triangleq \mathbf{G}\mathbf{X}_i$. Note that $\mathbf \Gamma \in \mathbb {C}^{M_i \times M_i}$ is a  diagonal matrix, whose $m$th diagonal entry $\Gamma(m)$ is the $m$th eigenvalue of $\mathbf{\Phi}_{M_i}$, and corresponds to the root Zadoff-Chu sequences as~\cite{ouyang2016orthogonal}
\begin{align} \label{eq:gamma}
    \Gamma(m)=e^{-\mathrm j \frac{\pi}{M_i}m^2}, \ \forall m, \ M_i \equiv 0 \ (\textrm{mod} 2).
\end{align}

The circulant property of discrete Fresnel transform (DFnT) \eqref{eq:ocdm eigen} allows the OCDM modulator to be integrated with a conventional OFDM modulator, using an additional DFT-based precoding operation $\mathbf{G}$. At the Tx, the data symbols are first mapped into $M_i$ sub-carriers and transformed to frequency domain as in \eqref{eq:ocdm eigen}. The resulting symbols are then serialized, passed through a pulse shaping filter, up-converted and transmitted. Define $T_i=\frac{1}{\Delta f_i}$ as the OCDM symbol duration. The time-domain OCDM signal can be written as
\begin{align}
\label{eq:ocdm time domain}
s_i(t) =\sum\limits_{n=0}^{N_i-1} \sum\limits_{m=0}^{M_i-1} [\mathbf X_{i}]_{m,n} e^{\mathrm{j}\frac{\pi}{4}}
e^{\frac{-\mathrm{j}\pi M_i}{T_i^2}\left(t-nT_i-\frac{mT_i}{M_i}\right)^2} \nonumber \\ e^{\mathrm j 2 \pi f_{c_i} t} \textrm{rect}(t-nT_i),
\end{align}
\normalsize
where $\textrm{rect}(t)\triangleq \left\{\def\arraystretch{1.2}\begin{tabular}{@{}l@{\quad}l@{}}
  $1$ &\; $0 \le t \le T$ \\
  $0$ & \;\textrm{otherwise}
\end{tabular}\right.$.
The Rx receives the radar return  \eqref{eq:ocdm time domain} over a doubly spread THz  radar channel \eqref{eq:generic channel} as the sum of reflections for each antenna output at frequency $f_{c_i}$, characterized by delay and Doppler shifts of the targets as follows
\par\noindent\small
\begin{align} 
\label{eq:ocdm rad output}
y_i^{\text{rad}}(t)\!&= \!\sqrt{\!\text{ PL}_{\text{LoS}}(f_{c_i},r_p)}\sum \limits_{p=1}^{P} \! h_p s_i(t-\tau_p) e^{j2\pi f_{c_i} \! \vartheta_{p}(t-\tau_p)} \!+\! w_i(t),
\end{align}
\normalsize
where $\vartheta_p=\frac{v_p}{c}$ is the normalized velocity and $w_i(t)\sim \mathcal{CN}(0,\epsilon_{i}^2)$ represents the additive white Gaussian noise (AWGN). Here, we assume $\frac{1}{M_iN_i}\sum \limits_{n=0}^{N_i-1} \sum \limits_{m=0}^{M_i-1} \mathbb E[|[\mathbf X_{i}]_{m,n}|]^2 \leq P_{\text{avg}}$, and thus the signal satisfies an average power constraint. Also, the subcarrier spacing is set to be larger than the maximum Doppler shift to maintain orthogonality, i.e., $f_{D_i}^{\max} \ll \Delta f_i$, $\forall i$, and thus $\Delta f_i$ is chosen based on the maximum velocity, such that this inequality is satisfied.  \textcolor{black}{In order to achieve unambiguous radar sensing at longer distances, waveform design without cyclic prefix is recommended~\cite{sturm2011waveform}. Hence, we consider a fixed phase constellation of payload data symbols that are modulated onto the chirps, which can be later nullified at the receiver.
Further, to avoid any ambiguity in distinguishing targets due to aliasing, the maximum delay spread of the targets should be less than the symbol duration within each subband, i.e., $\Delta \tau^{\max} < \min\limits_i{T_i}$ \cite{braun2014ofdm}.}
\section{Combined Target Estimation}
\begin{figure}[t]
    \centering
    \includegraphics[width=0.6\textwidth]{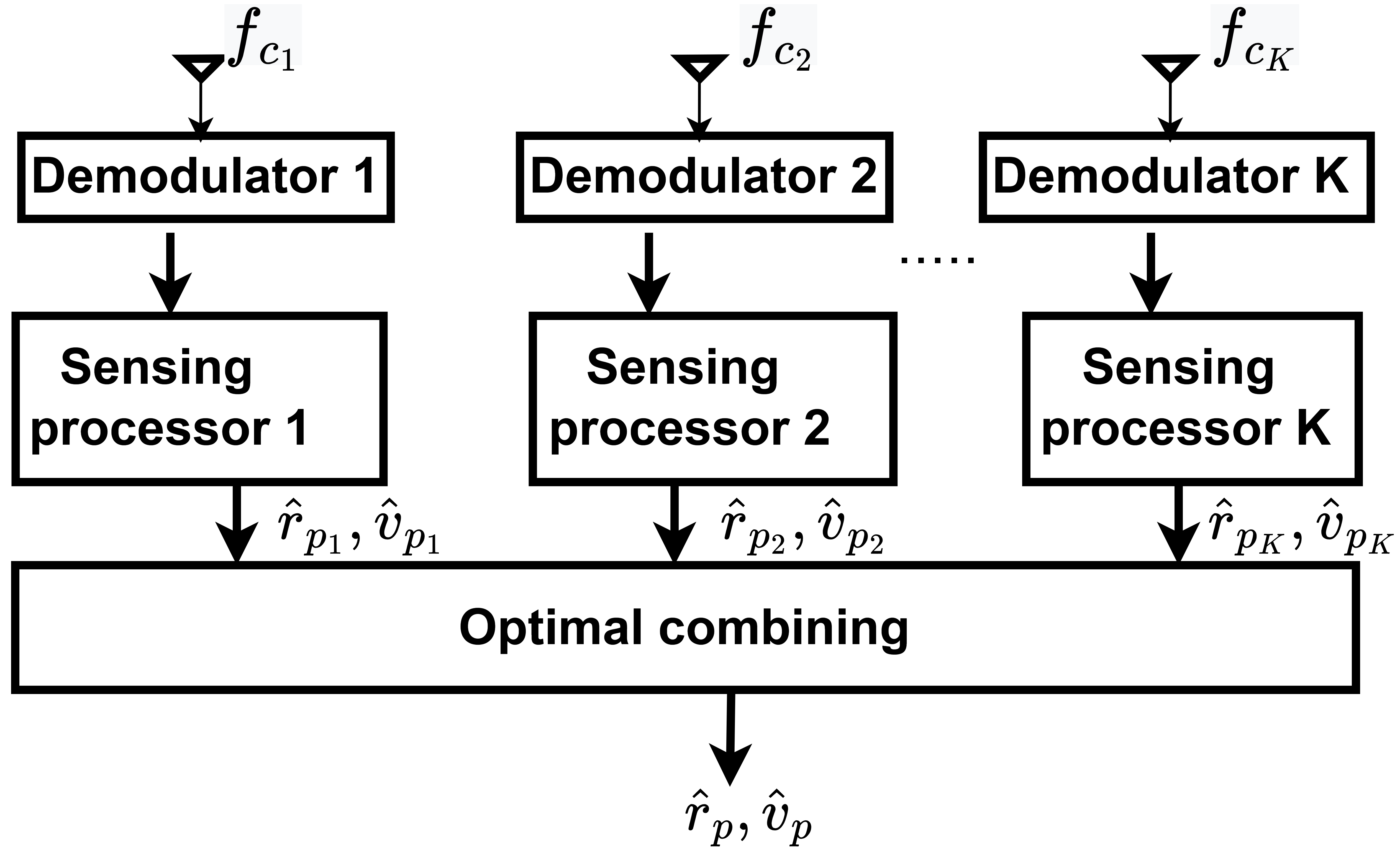}
    \caption{The receiver processing requires combining the estimates by assigning optimal weights to each subband.
    }
    \label{fig:Receiever}
\end{figure}
\textcolor{black}{Each demodulator $i$ tuned to center frequency $f_{c_i}$ (Fig. \ref{fig:Receiever}) receives the radar signal \eqref{eq:ocdm rad output}. After down conversion, the signal is sampled at $t=nT_i+m\frac{T_i}{M_i}$ to yield}
\par\noindent\small
\begin{align} \label{eq:radar samples}
&[\mathbf{Y}^{\text{rad}}_{i}]_{m,n}= \sum_{p=1}^{P} \tilde{h}_{p_i}  e^{\mathrm j2\pi \frac{f_{c_i}}{c}v_{p} (nT_i+m\frac{T_i}{M_i})}  \sum\limits_{m'=0}^{M_i-1} [\mathbf X_i]_{n,m'} e^{\mathrm j\frac{\pi}{4}} \nonumber \\ 
& \hspace{1.8cm} \times e^{-\mathrm j\pi\frac{M_i}{T_i^2}\left\lbrace(m-m')\frac{T_i}{M_i}-\tau_p\right\rbrace^2}+[\mathbf{W}_i]_{m,n}.
\end{align}
\normalsize
where $\tilde h_{p_i}=\sqrt{\text{ PL}_{\text{LoS}}(f_{c_i}, r_p)}h_p$. Next, we apply DFnT to observe the radar return across the chirps as follows 
\par\noindent\small
\begin{align} \label{eq: radar matched filtering}
&[\mathbf{Y}_i^{\text{rad}}]_{m,n} = \frac{1}{M_i} \sum \limits_{l=0}^{M_i-1} [\mathbf{Y}_i^{\text{rad}}]_{l,n}\exp (-\mathrm j\frac{\pi}{4}) \exp\left[\mathrm j \frac{\pi}{M_i}(m-l)^2\right] \nonumber \\
& \approx \!\sum \limits_{p=1}^{P} [\mathbf{X}_i]_{m,n} \tilde{h}_{p_i} e^{-\mathrm j\pi M_i \left(\frac{\tau_p}{T_i^2}\right)^2} \!\!e^{\mathrm j 2\pi \left (n \vartheta_p f_{c_i} T_i - m\tau_p\Delta f_i\right)}  \! +  [\mathbf{W}_i]_{m,n},
\end{align}
\normalsize
where the approximation follows because  $f_{D_i}^{\max} \ll \Delta f_i $. Since the payload data $\mathbf X_i$ is known at the radar receiver, we remove them from \eqref{eq: radar matched filtering} by performing an element-wise division. Note that the noise statistics do not change due to this operation \cite{braun2014ofdm}. Subsequently, the radar observations are (dropping the constant phase term)
\par\noindent\small
\begin{align} 
[\mathbf{Z}_i^{\text{rad}}]_{m,n}=\sum \limits_{p=1}^{P} \tilde{h}_{p_i}  e^{\mathrm j 2\pi \left (n \vartheta_p f_{c_i} T_i - m\tau_p\Delta f_i\right)}  +  [\mathbf{W}_i]_{m,n}.
\end{align}
\normalsize
\textcolor{black}{These observation samples are then fed to a sensing processor (SP) corresponding to each demodulator output, which outputs target parameter estimates.}
Assuming the number of targets has been determined (e.g., via hypothesis testing \cite{mishra2019sub}
we find maximum likelihood (ML) estimates of the target parameters $\boldsymbol{\theta}_i=[\theta_{1_i}, \cdots, \theta_{P_i}]^T$ such that $\theta_{p_i}=\left(r_{p_i}, v_{p_i}\right)$, $\forall p$, $\forall i$. We index the \emph{estimates} $r_{p_i}, v_{p_i}$ by the subband index $i$ because a separate estimate is obtained at each subband. The ground truth range and velocity are represented by $r_p$ and $v_p$, respectively.
The simplified log-likelihood function $\forall p$ is
\begin{align} \label{eq:radar estimate}
\begin{split}
 \scalebox{0.95}{$\mathscr{L}(\mathbf{Z}_i^{\text{rad}};\theta_{p_i})= 2 \tilde{h}_{p_i} \mathbb{Re}\left[\sum \limits_{m,n} 
 [\mathbf Z_i^{\text{rad}}]_{m,n} e^{-j2\pi n\vartheta_{p} f_{c_i}T_i}e^{ j2\pi m \tau_p\Delta f_i} \right]-\tilde h_{p_i}^2$}.
 \end{split}
\end{align}
Clearly, the parameters to be estimated in $\boldsymbol{\theta}_p$, $\forall p$ are decoupled. 
Note that the equation \eqref{eq:radar estimate} is a two-dimensional (2D) complex periodogram. The periodogram can be calculated by quantizing the frequencies and taking FFTs along the desired dimension \cite{braun2014ofdm}. By discretizing \eqref{eq:radar estimate}, the log-likelihood function is 
\begin{align} \label{eq:periodgram}
\mathscr{L}_q(\mathbf{Z}_i^{\text{rad}}; m',n')=2 \tilde h_{p_i} \mathbb {R}\left[\sum \limits_{m=0}^{M_{\text{Per}}-1} \sum \limits_{n=0}^{N_{\text{Per}}-1} [\mathbf Z_i^{\text{rad}}]_{m,n} e^{-j2\pi \frac{nn'}{N_{\text{Per}}}}e^{ j2\pi \frac{m m'}{   M_{\text{Per}}}}\right],
 \end{align}
\textcolor{black}{where $m'\coloneqq \tau_p \Delta f_i$ and $n'\coloneqq \vartheta_pf_{c_i}T_i$ are the discretized frequencies in \eqref{eq:radar estimate} over the search grid $m'=0, \cdots, M_{\text{Per}}-1$ and $n'=-\frac{N_{\text{Per}}}{2}, \cdots, \frac{N_{\text{Per}}}{2}-1$, with $M_{\text{Per}}>M_i$ and $N_{\text{Per}}>N_i$, $\forall i$.} Note that \eqref{eq:periodgram} is a 2D-DFT applied to $\mathbf{Z}_i^{\text{rad}} \in \mathbb{C}^{M \times N}$ with oversampling in both dimensions. The ML solution of $\theta_{p_i}$ is 
\begin{align}
   [\hat{m'}, \hat{n'}]=\operatorname{\text{argmax}}_{{m' \in \mathcal{M},n' \in \mathcal N}}  \mathscr{L}_q(\mathbf{Z}_i^{\text{rad}};m',n')
\end{align}
Thus, we obtain peaks at $m_p'$th and $n_p'$th bin of \eqref{eq:periodgram}, which correspond to the delay and Doppler, respectively, at each 
SP $i$ as follows
\begin{align}
\label{eq: estimated delay dopp}
\widehat{\tau}_{p_i} =  \frac{\hat{m'}}{\Delta f_i M_{\text{Per}}}, \ \widehat{\vartheta}_{p_i} =\frac{n'}{2 \pi f_{c_i}T_i N_{\text{Per}}}
\end{align}
The range and velocity of the target $(r_{p_i},v_{p_i})$, $\forall p$ are thus estimated. 
\begin{figure*} \includegraphics[width=1.0\textwidth]{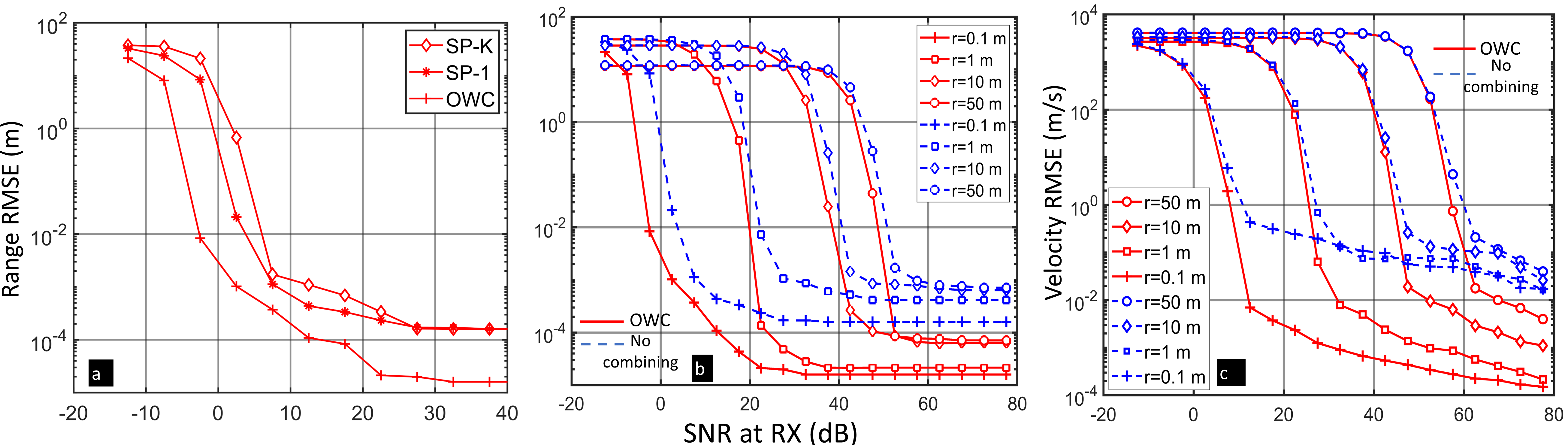} 
 \caption{(a) Effect of frequency-dependent THz path loss on range estimation versus the SNR, with $r=0.1$ m; effect of distance and frequency-dependent THz path loss on (b) range estimation (c) velocity estimation versus the SNR.}
 \label{fig:estimation}
\end{figure*}

In our optimal weighted combining (OWC), the estimates from each of the $K$ sensing processors need to be combined to get a final estimate of the target range and velocity $\theta_p=(r_p,v_p)$, $\forall p$. Without loss of generality, we present the combining scheme for a single target scenario. Hence, we drop the target subscript $p$ in the subsequent analysis. The estimated parameters of the target at each demodulator block $i$ can be linearized under a first order Taylor series approximation, considering small errors, as
\begin{align}
\hat{\boldsymbol{\zeta}}=\mathbf{1}_K\zeta+\mathbf{e}_{\boldsymbol{\zeta}} \ \in \mathbb{C}^{K}.
\end{align}
where $\hat{\boldsymbol{\zeta}}=[\hat \zeta_1, \cdots, \hat \zeta_K]^T$ with $\zeta_i \in \{ \hat r_i, \hat v_i \} $ being the estimated parameter at SP $i$, 
$\zeta\in \{ r,v \}$ is the true target parameter and $\mathbf{e}_{\boldsymbol{\zeta}}=[e_{\zeta_1}, \cdots, e_{\zeta_K}]^T$ is the estimation error. At high SNR, the DFT based estimator achieves the Cram\'{e}r-Rao lower bound (CRLB). Thus, we can approximate the variance of $\mathbf{e}_{\boldsymbol{\zeta}}$ at high SNR  as  \cite{sang2022integrated}:
\begin{align} \label{eq:CRLB}
    \begin{split}
        \sigma_{r_i}^2 \approx \frac{6  \epsilon_i^2}{\left(2\pi)^2M_iN_i(N_i^2-1\right) |\tilde h_i|^2 P_{\text{avg}}} \left(\frac{c}{\Delta f_i}\right)^2\\
        \sigma_{v_i}^2 \approx \frac{6 \epsilon_i^2}{(2\pi)^2M_iN_i(M_i^2-1) |\tilde h_i|^2 P_{\text{avg}}} \left(\frac{c}{T_i f_{c_i}}\right)^2.
    \end{split}
\end{align}
\setlength{\textfloatsep}{0pt}
\begin{proposition}
\label{prop: sufficient stats}
Define $T(\hat{\boldsymbol\zeta})=\boldsymbol\beta^T \hat {\boldsymbol \zeta}$; then $T(\hat{\boldsymbol\zeta})$ is a sufficient statistic for estimating $\boldsymbol \zeta$, where $\boldsymbol \beta=[\beta_1, \cdots, \beta_K]^T \in \mathbb R^K$ denotes the weights of the linear combiner.
\end{proposition}
\noindent The proposition follows from the fact that error at each SP 
is independent and hence $T(\hat{\boldsymbol\zeta})$ can be modelled as being corrupted by AWGN using the central limit theorem when there are a large number of estimates. Then, it is easy to show that $T(\hat{\boldsymbol\zeta})$ can be factorized to satisfy Neymen Fisher factorization theorem \cite{kay1993fundamentals}.
\begin{theorem}
\label{thm: optimization}
The optimal combining scheme is a linear weighted combination of the estimates obtained from the $K$ sensing processors which minimizes the estimation error, i.e.,
\begin{align}
\label{eq: optimization}
&\min \limits_{\boldsymbol \beta} \mathbb{E} \left \lbrace \left(\boldsymbol \beta^H \hat{\boldsymbol \zeta}-\zeta \right)^2 \right \rbrace \text{ s.t. } \  \mathbf 1_K^H \boldsymbol \beta=1,
\end{align}
\noindent and the optimal combining weights are given by
\begin{align}
\label{eq:opt weight}
    \beta_k=\frac{\sigma_{\zeta_k}^{-2}}{\sum_{i=1}^{K} \sigma_{\zeta_i}^{-2}}, \ \forall k \in \{ 1, \cdots, K\}.
\end{align}
\end{theorem}
\begin{IEEEproof}
The optimization problem in \eqref{eq: optimization} can be simplified as 
$\min \limits_{\beta_1, \cdots, \beta_k} \left[ \boldsymbol{\beta}^H \mathbf{R}_{e_\zeta} \boldsymbol{\beta}\right]$  subject to 
$\mathbf{1}_K^H\boldsymbol{\beta}=1,$
where $\mathbf R_{e_\zeta}=\mathbb{E}[\mathbf e_{\zeta} \mathbf e_{\zeta}^H]$ is the error covariance matrix, with $\sigma_{\zeta_i}^2$, $\forall i$ \eqref{eq:CRLB} its diagonal elements. The Lagrangian of this optimization problem is
    $\mathcal{L}=\boldsymbol \beta^H \mathbf{R}_{e_{\zeta}} \boldsymbol \beta + \lambda (\mathbf{1}^H_K \boldsymbol \beta -1),$
where $\lambda$ is the Lagrange parameter.
Taking the derivatives with respect to $\boldsymbol \beta$ and $\lambda$, setting it to zero, and simplifying, we get
 $\boldsymbol \beta=-\frac{\lambda}{2}\mathbf{R}_{e_{\zeta}}^{-1}\mathbf{1}_K$ and $\lambda=-\frac{2}{\mathbf 1_K^H\mathbf R_{e_\zeta} \mathbf{1}_K}$, which results in the optimal weights given by \eqref{eq:opt weight}.
\end{IEEEproof}
\section{Numerical Experiments}
We evaluate the performance of the proposed THz MCW-OCDM system through numerical experiments. Throughout all experiments, the total bandwidth $B$ is set to $1.4$ THz. Based on the available distance-dependent transmission windows \cite{hossain2019hierarchical}, we divide the total bandwidth into subbands $B_i$, each spanning $1$ GHz. We consider equal number of chirps and OCDM symbols, i.e., $M=256$ and $N=256$, for each subband. Thus, the spacing $\Delta f_i$ between two chirps at each subband is 3.9 MHz, which is well within the coherence bandwidth of the THz channel \cite{sarieddeen2021overview}. The OCDM frame time $T$ is $0.25~\mu$s. We place a reference target at different distances from the radar transceiver, and with velocity of $23$ m/s. The noise variance $\epsilon_i^2$ is 1 and the THz path loss at each $f_{c_i}$ for different $r$ is given by \eqref{eq:path loss} \cite{hossain2019hierarchical}.  \textcolor{black}{We set the Tx power such that at $r=0.1$ m and for SP-1, the received SNR is varied from $-12$ dB to $15$ dB. This is the reference Rx SNR for all the plots at all distances.}
We use the metric root mean square error (RMSE) $= \sqrt{\mathbb{E}\left\lbrace||\hat{\boldsymbol \zeta }-\zeta||^2\right\rbrace}$ to quantify the estimation accuracy.  

Fig. \ref{fig:estimation}a compares the RMSE of range using our proposed OWC scheme with that of individual SP estimates at THz frequencies as a function of the SNR, when the target is at a distance of $r=0.1$~m. Note that, at this distance, radar returns are obtained at all bandwidth windows \cite{hossain2019hierarchical} and OWC yields sub-millimeter level sensing accuracy at reasonable Rx SNR levels ($> 0$ dB). Further, at RMSE of $10^{-3}$, OWC outperforms the estimates from SPs at the first and last bandwidth windows by 6 dB approximately. This is also the case when the SP estimates are not combined and only the average RMSE is considered. Furthermore,  Fig. \ref{fig:estimation}b shows the impact of target distance on RMSE as a function of SNR. Clearly, the OWC significantly outperforms the estimates of a single SP at all distances and the improvement is more pronounced at lower distances. This is because, by assigning optimal weights, the combiner is able to tackle the THz band path loss much better than the single estimate at SP-1. Further, a higher number of SPs are available at lower distances, resulting in better target localization.

Fig. \ref{fig:estimation}c highlights the OWC advantage in velocity estimation 
for different bistatic target distances $r$. The OWC improves the RMSE by nearly $6$ dB for all target ranges. As the target range increases, radar returns are obtained from smaller number of bandwidth windows and the path loss also becomes more severe. The increase of path loss with frequency also results in the worsening of the velocity estimates across THz subbands, 
which, in turn, increases the estimation error. 
\section{Summary}
We presented a multi-wideband OCDM framework to overcome the limitation of both frequency and distance dependent path losses at THz band in radar target parameter estimation. We provided a novel multi-stage sensing algorithm to make use of data frames from different THz subbands. 
To this end, we derived optimal combining weights across the different sensing processors based on the Cram\'{e}r-Rao lower bound on the parameter estimates. We demonstrated that using an optimal weighted combiner for processing the radar returns from different THz transmission window significantly enhances the estimation accuracy by prioritizing the more accurate estimates from subbands which experience lower absorption losses. 
\bibliographystyle{IEEEtran}
\bibliography{ref.bib}	
\end{document}